\documentstyle[preprint,aps]{revtex}
\begin{document}
\draft
\title{Raman Scattering from Fractals. Simulation on Large
Structures by the Method of Moments}
\author{G.Viliani, R.Dell'Anna, O.Pilla and M.Montagna}
\address{INFM, Dipartimento di Fisica, Universit\'a di Trento, I-38050 Povo,
Trento, Italy}
\author{G. Ruocco and G.Signorelli}
\address{INFM, Dipartimento di Fisica, Universit\'a dell'Aquila, Coppito,
I-67100 L'Aquila, Italy}
\author{V.Mazzacurati}
\address{INFM, Dipartimento di Scienze e Tecnologie Biomediche e
Biometria (STBB),\\
Universit\'a dell'Aquila, Collemaggio, I-67100 L'Aquila, Italy }
\date{\today}
\maketitle
\begin{abstract}
We have employed the method of spectral moments to study the density of
vibrational states and the Raman coupling coefficient of large 2- and 3-
dimensional percolators at threshold and at higher concentration. We first
discuss the over-and under-flow problems of the procedure which
arise when -like in the present case- it is necessary to calculate
a few thousand moments. Then we report on the numerical results;
these show that different scattering mechanisms, all {\it a priori} equally
probable in real systems, produce largely different coupling coefficients
with different frequency dependence. Our results are compared with existing
scaling theories of Raman scattering. The situation that emerges is
complex; on the one hand, there is indication that the existing theory is
not satisfactory; on the other hand, the simulations above threshold
show that in this case the coupling coefficients have very little resemblance,
if any, with the same quantities at threshold.
\end{abstract}
\section{Introduction}
\label{intro}
Crystalline solids are characterised by translational invariance and this
symmetry property makes it possible to derive explicit analytic expressions
for the vibrational eigenvectors in the harmonic approximation or, at worst,
entails the diagonalization of small matrices. As a
consequence, a lot is known on many physical properties that are
determined by the vibrational dynamics of crystals (vibrational density of
states, thermal properties, inelastic light and neutron scattering, and so
on). The situation is completely different in disordered solids: in this
case the lack of translational invariance often requires that the dynamical
properties be studied by numerical calculations on more-or-less realistic
models.
\par
Even when the model well represents the real system, one of the factors that
make a numerical calculation useful is the dimension of
the sample; in many cases, the bigger the dimension, the lower the
probability that finite size effects dominate the results and make any
comparison with experimental data problematic.
\par
As regards the study of the vibrational dynamics of disordered systems, and
assuming that the harmonic approximation holds, one possible numerical approach
is to
build up the dynamical matrix and diagonalize it; this provides vibrational
eigenvalues and eigenvectors, i.e. all the information that may be
required. The problem is that the linear size of the matrix to be
diagonalized is as large as the number of vibrational degrees of freedom,
and, for example, a sample containing  $10 \times 10 \times 10$ atoms would
require the diagonalization of a $3000 \times 3000$ matrix: a very large
matrix for not too large a system. Even though in some cases sufficient
insight may be gained by considering only one degree of freedom per atom,
the calculations are in general lengthy and expensive.
In any case, 3-dimensional
models containing, say, 20$\times$20$\times$20 atoms are at present
practically intractable with this technique.
\par
In many cases all the detailed information contained in the
eigenvalues and eigenvectors is not necessary. For example, the density of
vibrational states is a more significant quantity to compare with
experiment than the sequence of eigenvalues. The important thing is that
some of these relevant quantities can be calculated without diagonalizing
the dynamical matrix, by exploiting in a different and more efficient
way the information it contains.
\par
In the present paper we will use the method of moments
\cite{1benoit1,1Abenoit2}
to compute the density of vibrational states and the Raman coupling
coefficient of model disordered systems consisting of two- and
three-dimensional (hereafter 2D and 3D respectively) site and bond
percolators; these were studied both at percolation threshold concentration
and at higher concentration.
Percolators at threshold are fractals and using arguments based on scale
invariance it has been shown \cite{2alexorb,3rammtoul} that the density of
vibrational states follows a power law of the type $\rho(\omega) \propto
\omega^{\overline d-1}$, where $\overline d$ is a parameter known as the
spectral dimension.
\par
In the recent literature several attempts were made to find a similar power
law for the Raman
coupling coefficient $C(\omega)$, which determines the (Stokes)
scattered intensity in the following way:
\begin{equation}
	I(\omega) = (n(\omega)+1) C(\omega) \rho(\omega)/\omega
\label{Iom}
\end{equation}
where $n(\omega)$ is the Bose-Einstein population factor. As we will discuss in
the following sections, the proposed scaling laws $C(\omega) \propto \omega^x$
are
based
on assumptions that are not
universally accepted so that comparison with numerical simulation is of great
importance. Numerical calculations of $C(\omega)$ were earlier
produced for a number of
percolators at threshold \cite{4lettera,5colori,6ericstoll} using dynamical
matrix diagonalization. As mentioned, this allows for only limited
dimensions of the models, and this restriction is particularly bad in this
case, since there are indications \cite{6ericstoll,7orbnippi} that power
laws for $C(\omega)$
might apply only at low frequency. In the present paper we present
an exhaustive analysis of $C(\omega)$ in much larger systems, in order to avoid
as far as possible finite size effects and reach the lowest possible
frequency.
\par
In section II we summarize the method of calculation; section III
is devoted to
the stability and precision problems of the procedure and to the way we were
able to overcome over- and under-flow difficulties; in section IV we present
the numerical data on various types of percolators; the data are discussed
in connection with the phenomenology of Raman scattering from real disordered
solids in section V.

\section{The method of moments}
\label{method}
In the following we shall outline the method of spectral moments, which is
reported in detail in the papers by Benoit and coworkers \cite{1benoit1}.
This method applies to the case where one is interested in calculating a
spectral function $f(\omega^2)$ of the form:
\begin{equation}
	f(\omega^2)=\sum_{\lambda=1}^{3N}d_\lambda^2\delta(\omega^2-\omega_\lambda^2)
\label{fo2}
\end{equation}
where
\begin{equation}
	d_\lambda=\sum_{i=1}^{3N}p_i e_i(\lambda)
\label{dl}
\end{equation}
The system consists of N
identical masses which interact via identical harmonic potentials and is
characterized by the $3N \times 3N$ dynamical matrix ${\bf D}$ with eigenvalues
$\omega_\lambda$ and eigenvectors $e_i(\lambda)$; $\lambda=1...3N$ labels
the normal
modes and $i$ is a
collective index
that labels the masses ($l=1...N$) and the cartesian components
($\alpha=x,y,z$) of the eigenvectors, $i=(l,\alpha$).
The coefficients $p_i$ depend on the spectral function to be computed; the
explicit expressions of $p_i$ for the cases of interest in this paper are
discussed in the Appendix.
\par
It has been shown by Benoit and coworkers \cite{1benoit1} that $f(\omega^2)$
can
be put in the form
\begin{equation}
f(\omega^2) = -\frac{1}{\pi} \lim_{\varepsilon \rightarrow 0^+}
{\rm{Im}\{ R(z)\}}
\label{fo2a}
\end{equation}
where $z=\omega^2+i\varepsilon$, and
$$
R(z) = \int_{-\infty}^{+\infty}{\frac{f(\omega^2)}{z -\omega^2}d\omega^2}
$$
The latter can be developed as a continuous fraction
\begin{equation}
R(z)=
\frac{1}{z-a_1- \displaystyle{
\frac{b_1}{z-a_2- \displaystyle{
\frac{b_2}{z-a_3-......}}}}}
\label{Rz}
\end{equation}
where the real coefficients $a_n$ and $b_n$ depend on the generalized
moments $\nu_{nm}$ and $\overline \nu_{nm}$ of $f(\omega^2)$ (whence the
name of the method):
\begin{equation}
a_{n+1} = \frac{\overline \nu_{nn}}{\nu_{nn}},\;\;\;\;\;\;\;
b_n = \frac{\nu_{nn}}{\nu_{n-1,n-1}}
\label{anbn}
\end{equation}
where
\begin{equation}
  \nu_{nn}= \int_0^{1}P_n(\omega^2)P_n(\omega^2) f(\omega^2) d\omega^2
\label{nu}
\end{equation}
and
\begin{equation}
 \overline \nu_{nn}= \int_0^{1}P_n(\omega^2)P_n(\omega^2) f(\omega^2)
\omega^2\; d\omega^2
\label{vub}.
\end{equation}
$\{P_n(\omega^2)\}$ is the succession of polynomials orthogonal  with
respect to the spectral function $f(\omega^2)$ (as in the rest of this
paper, the frequency scale in equations 7 and 8 has been normalized so that the
maximum frequency is 1).
The moments in turn can be computed by a recursive procedure that makes use
of the dynamical matrix. Let ${\bf t}^{0}$ be the normalized
vector with components proportional to $p_i$, it
turns out that \cite{1benoit1}
\begin{equation}
\nu_{nn} = ({\bf t}^{(n)}, {\bf t}^{(n)}),\;\;\;\;\;\;
\overline\nu_{nn} = ({\bf t}^{(n)}, {\bf D}{\bf t}^{(n)})
\label{nunub}
\end{equation}
where ${\bf t}^{(n)}$ is a vector which obeys the recursive relation
\begin{equation}
{\bf t}^{(n+1)} = ({\bf D} - a_{n+1}) {\bf t}^{(n)}
- b_n {\bf t}^{(n-1)}
\label{tn}
\end{equation}
Equations \ref{anbn} to \ref{tn}, with ${\bf t}^{(-1)} = 0$,
determine $a_n$ and $b_n$ recursively, and these give $f(\omega^2)$ through
$R(z)$.
\par
In principle, if one computed a number of $a_n$ and $b_n$
coefficients equal to the number of distinct eigenvalues \cite{1benoit1},
in the limit $\varepsilon \rightarrow 0$ one would reproduce exactly the
response
function in Eq. (\ref{fo2a}).
\par
In
computations of practical interest (i.e. on systems with a large number of
masses) this is neither possible nor necessary. The number of moments to be
computed and the value of $\varepsilon$ depend
on the frequency
range where  the spectral function is to be reproduced most carefully.
\par
The use a finite value for $\varepsilon$, which is equivalent to
considering
the convolution of $f(\omega^2)$ with a lorentzian
$$
L_\varepsilon(\omega^2) = \frac{\varepsilon}{\pi(\varepsilon^2+ \omega^4)},
$$
makes it
impossible to resolve spectral features with width of the order of (or
smaller than) $\varepsilon$.
\par
The fact that only a limited number of moments is computed introduces a
truncation error that, as noted by Turchi et al. \cite{8turchi},
can be minimized when the spectral
function $f(\omega^2)$ has
no gaps and no divergences. In this case the coefficients $a_n$ and $b_n$
tend to make
small fluctuations about constant asymptotic values as $n \rightarrow
\infty$, and this suggests to substitute the asymptotic values in Eq.
(\ref{Rz})
for the
coefficients that are not computed. If one wants only $M$ moments one
re-writes Eq. (\ref{Rz}) as:
$$
R(z) = \frac{1}{z-a_1 \displaystyle{- \frac{b_1}{z-a_2 \cdot
\cdot \cdot - \displaystyle{ \frac{b_{M-1}}{z-a_M-T_M(z)}}}}}.
$$
and under the assumption of constant $a_{n}$=$a$ and $b_{n}$=$b$ for $n \geq
M$,
$T_M(z)$
is easily found to obey
$$
	T_M(z) = \frac{1}{z-a-bT_M(z)}.
$$
Since in our case $a_n$ and $b_n$ continue to make small
oscillations about $a$ and $b$ respectively for large $n$,
this procedure provides only an approximation to the true $R(z)$;
in any case $T_M(z)$ becomes less and less important as the number of
moments
that are actually computed increases.

\section{Stability and precision of the procedure}
\label{stabi}
Since we will be mainly interested in studying the behavior of spectral
functions
like in Eq. (\ref{fo2}) at frequencies of the
order of $10^{-3} \div 10^{-2}$ times
the Debye frequency, the use of many (some thousand) moments is necessary,
as can
be seen empirically and deduced theoretically. This requires many iterations
of Eq. (\ref{tn}) that can introduce numerical errors in the computed
$f(\omega^2)$. The errors can arise both from the precision of the calculation
(accumulation of roundoff approximations) and/or from the instability of the
algorithm, i.e. from the sensitivity of the algorithm to small random
departures of $a_n$ and $b_n$ from the true values.
\par
As a first example in Fig. 1 we show the effect of the number of
computed moments on the density of states of 2D,
650$\times$650 site percolators at percolation threshold, consisting of
identical
masses connected by identical springs; each mass has only one degree of
freedom, cyclical boundary conditions are imposed \cite{9Aroyer}.
It is clearly seen that
in passing from 50 to 5000 iterations the minimum frequency at which the
expected density of states ($\rho(\omega) \propto \omega^{0.31})$
\cite{stauffer}
is reproduced, decreases by more than one order of magnitude. Note
that 1000 iterations are not sufficient. The origin of the dip at
$\omega \approx 0.7$ will be discussed later.
\par
The necessity of so many iterations may be justified theoretically as
follows. Using the orthogonality relation
$$
  \int_0^{1} P_n(\omega^2)\omega^{2k}f(\omega^2)d\omega^2=0
$$
with $k=0,...,n-1$, equation (\ref{nu}) becomes:
$$
 \nu_{nn}= \int_0^{1} P_n(\omega^2)(\omega^2-1/2)^n f(\omega^2) d\omega^2
$$
from which it is easily seen that, for large $n$, $\nu_{nn} $ is mostly
determined by the extremes of the $\omega$ range, so that the new information
that is added at each step is more and more concerned with the extremes;
the new information is necessary to approach the minimum attainable
frequency, as seen in Fig. 1 \cite{10minfreq}. Moreover, it is also clear
that in
order to reproduce the behaviour of $f(\omega^2)$ in the central region of
frequencies it is neither necessary nor (as we will see) advisable to use
many iterations.
\par
The first numerical difficulty that is met in the calculation of many
moments is that $\nu_{nn}$ (and $\overline \nu_{nn}$) tend to approach $0$ or
$\infty$, causing overflow or underflow of the
computed moments. This difficulty can be overcome by noting
\cite{1benoit1} that if one scales the physical characteristics of the
system (masses and/or elastic constants) by a factor $s$, the generalized
 moments scale as well:
\begin{eqnarray}
\label{scale}
	{\cal K} &\rightarrow& s{\cal K} \nonumber \\
    \nu_{nn} &\rightarrow&  s^{2n}\nu_{nn} \nonumber\\
    \overline \nu_{nn} &\rightarrow& s^{2n+1} \overline \nu_{nn} \nonumber\\
    b_n &\rightarrow& s^2 b_n \nonumber\\
	a_n &\rightarrow& s^2 a_n
\end{eqnarray}
It is therefore possible to choose a value of $s$ such that the generalized
moments are finite. Since $b_n=\nu_{nn} /\nu_{n-1,n-1}$, one possibility is
to start with an arbitrary elastic constant ${\cal K}_1$, then compute a
sequence of
$M_1$ coefficients $b_n$ until $\nu_{M_1M_1} $ reaches the
over/under-flow condition, then
compute $\langle b\rangle =M_1^{-1}\sum_nb_n$ and scale ${\cal K}_1$
in agreement with the first of equations \ref{scale} so
that $\langle
b\rangle $ is equal to a desired value.
\par
At first sight, it would seem that
$\langle b\rangle =1$
is a good choice to avoid divergences, so that for the scaling factor $s$ we
have
$$
s^2 = \frac{1}{\langle b\rangle}
$$
This procedure may be repeated several times, until the desired number of
moments can be calculated.
\par
However $\langle b \rangle =1$ is not the
most
convenient choice. In fact, since the $b_n$'s fluctuate around  $\langle
b\rangle$, we can put $b_n=\langle b \rangle (1+\varepsilon_n)$, with
$\langle \varepsilon_n\rangle = 0$; in this way we have
$$
\nu_{nn} =\langle b\rangle
(1+\varepsilon_n)\nu_{n-1,n-1}
$$
or
\begin{eqnarray}
\label{vu1}
	\nu_{nn}&=& \langle b \rangle^n \nu_{00}\prod_{i=1}^n (1+\varepsilon_i)
       \nonumber \\
       &=&\langle b \rangle^n \nu_{00}(1+\sum_i\varepsilon_i+\sum_{i \neq j}
     \varepsilon_i\varepsilon_j + ...)
\end{eqnarray}
Since $\sum_i\varepsilon_i =0$, we have
$$
\sum_{i\neq j} \varepsilon_i \varepsilon_j = \sum_{ij}\varepsilon_i
\varepsilon_j -
\sum_i\varepsilon_i^2 = -\sum_i\varepsilon_i^2 = -n\langle \varepsilon
^2\rangle
$$
so that
\begin{equation}
\nu_{nn} \approx \nu_{00}\langle b\rangle ^n(1-n\langle \varepsilon^2\rangle
) \approx \nu_{00}\langle b\rangle ^n e^{-
n\langle \varepsilon^2\rangle }
\label{vu2}
\end{equation}
 Thus, in order to control the divergence of $\nu_{nn} $ one should not
impose $\langle b\rangle =1$, but
\begin{equation}
\langle b\rangle =e^{\langle \varepsilon ^2\rangle }.
\label{b}
\end{equation}
We have verified that with this choice the number of corrections of
the elastic constant necessary for all the 5001 iterations of a
medium-sized system (square lattice of linear size 100), or for 5000
iterations of a larger system (2D percolator of linear size
650), is smaller than with the first choice.
\par
However, neither correction procedure eliminates a
further problem concerning the orthogonality of the series $\{{\bf t}^{(n)}\}$.
In fact, from Eq. (\ref{tn})  it is easy
to see that $({\bf t}^{(n)},{\bf t}^{(m)})\propto \delta_{nm}$, but in
practical calculations the orthogonality is lost if $|n-m|$ is large enough.
\par
In Fig. 2 we report $({\bf t}^{(0)}, {\bf t}^{(n)})$ as a function of $n$ for
four different systems: full (ordered) square lattices of linear sizes 50
and 100, and site percolators at threshold concentration of linear sizes 50
and 650. The calculations were carried out in double precision, but using
single precision only increases the value of the initial plateau. It is
interesting to note that the value of $n$
at which orthogonality begins to fade does not depend in a simple way on
the number of masses (and therefore on the number of required computer
operations): in fact, for the larger ordered lattice orthogonality is
maintained for a greater number of iterations, while for percolators this
effect is much less evident, if any.
\par
This behaviour can be understood by noting
that the recursive relation  \ref{tn}
is very similar to the one used in the Lanczos procedure of
tridiagonalization, and in fact the two become identical if in the Lanczos
procedure \cite{11golub} one makes the substitutions:
\begin{eqnarray*}
    {\bf r}_j &\rightarrow& {\bf t}_j/\sqrt{{\bf t}_j \cdot {\bf t}_j}\\
    \alpha &\rightarrow& a\\
    \beta^2 &\rightarrow& b
\end{eqnarray*}
As discussed at length in reference \cite{11golub}, this loss of
orthogonality is
inherent and not caused by accumulation of numerical roundoff errors.
Moreover, it is known that the Lanczos
procedure without any re-orthogonalization works well for the extreme
eigenvalues and eigenvectors, while there are problems for the inner range.
\par
{}From another point of view, the iterative equation (\ref{tn}) may be
considered as a non linear logistic application (the non linearity
coming from the dependence of $a_n$ and $b_n$ on ${\bf t}^{(n)}$) in an
$N$-dimensional space. It is known that these equations admit chaotic
solutions for the ${\bf t}^{(n)}$ variables, i.e. they produce sequences
$\{{\bf t}^{(n)}\}$ that, as $n$ grows, become very unstable with respect
to little variations of ${\bf t}^{(0)}$. Therefore, for
example, from ${\bf t}^{(0)}$ numerically we obtain
${\bf t}'^{(1)}= {\bf t}^{(1)} + \delta{\bf t}^{(1)}$,
$\delta{\bf t}^{(1)}$ being the machine precision error. Consequently, because
of the intrinsic chaoticity of equation (\ref{tn}), the computed
$\{{\bf t}'^{(n)}\}$
sequence will diverge from the "true" one and
$({\bf t}^{(0)},{\bf t}'^{(n)}) \neq 0$.
\par
Our numerical data are in complete agreement with the above results of
matrix calculus. First of all, when we compute the density of states (or the
Raman coupling coefficient) of a linear chain, orthogonality is
preserved no matter how long the chain; this is expected because in this
case the dynamical matrix is tridiagonal.
\par
Moreover, when comparing the
outcome of the method of moments with the eigenvalues of an exactly
solvable model (square harmonic lattice with identical masses
and springs with fixed
boundary conditions \cite{9bowers}) we find that as orthogonality begins to
be lost the peaks of the density of states no longer fit the exact
eigenvalues
in the central part of the (linear) spectrum, while the extreme eigenvalues are
still
well fitted, provided we use a sufficient number of moments. In any case,
the central part of the spectrum is perfectly fitted only for systems small
enough that computation of {\it all} moments does not result in
non-orthogonal vectors. For larger systems, increasing the number of moments
 worsens the look of the central part of the spectrum, be it the
density of states or the Raman coupling coefficient, and this is the origin
of the dip in the density of states of Fig. 1. In our case, this is not a
serious problem. In fact, on the one hand we are mostly interested in the
low frequency part of the coupling coefficients, this being the very reason why
we compute
so many moments; on the other hand, if we should be interested in the
central range we would compute only a few moments. This is possible
because the coupling coefficients are smooth functions due to disorder, we
don't
need to resolve one
peak from the next, all we need is a frequency-averaged result.
\section{Numerical results on Raman coupling coefficients}
\label{nume}
We proceed now to presenting the numerical data. As mentioned in the
introduction, the problem of finding a scaling law for the Raman coupling
coefficient in fractals, if it exists, has been the object of several
experimental and theoretical papers
\cite{6ericstoll,12bouk,12Aeurlet,12Bcarturan,13tsujimi,14physica}.
Besides being interesting on their own,
fractals are thought to be reasonably representative and relatively simple
models of important classes of disordered systems, especially as regards
vibrational characteristics. Raman scattering is one of the most convenient
experimental techniques to study vibrations in disordered solids: it does
not require big facilities and, contrary to the case of crystals, all
vibrational modes contribute to the scattering in a disordered system. The
price to be paid is that $C(\omega)$ in equation (\ref{Iom}) is not known
{\it a priori}, so
that, as mentioned, an important question regards whether or not it can be
cast in the form $C(\omega) \propto \omega^x$, and what expression  $x$ has in
terms
of the spectral dimension $\overline d$, of the fractal dimension $D$, and
possibly of other parameters which characterize the fractal.
\par
All the expressions proposed for $x$ so far were derived under the
assumption that $C(\omega)$ can be determined by considering the scaling
properties of the strain induced by the vibrations of the fractal (the
so-called fractons)
\cite{6ericstoll,12bouk,12Aeurlet,13tsujimi,14physica,16keyes,17aldo,19emilio}.
In two recent papers \cite{4lettera,5colori}
we argued against such possibility on the basis of numerical
calculations of $C(\omega)$ in 2D and 3D site percolators, which apparently
showed that ($i$) none of the proposed $x$ could fit the numerical data in
2D,
and ($ii$) a single $x$ seems not to be sufficient in 3D for the Dipole-
Induced-Dipole (DID) scattering mechanism \cite{5colori} (see Appendix).
\par
The conclusion we arrived at on the basis of these numerical results, i.e.
that the existence of scaling for $C(\omega)$ is not always evident in
fractals,
has been recently challenged on the basis of the following arguments: ($i$)
the size of the percolation clusters employed (60$\times$60 in 2D,
29$\times$29$\times$29 in 3D) was too small to allow positive conclusions to
be drawn on scaling, which is expected to hold at low frequency
\cite{7orbnippi}, and ($ii$) even more so for {\it site}-percolators: {\it
bond}-percolators should be used instead, because for the latter the scaling
regime starts at higher frequency \cite{6ericstoll}.
\par
These issues will be discussed in the following, after we have reported
numerical data obtained with the method of moments for a variety of much
larger systems including site- and bond-percolators at and above percolation
threshold, whose scattering units may or may not have random electrical
polarizabilities (electrical disorder), and assuming different scattering
mechanisms: DID, DID truncated to nearest neighbors (NNDID),
and Bond-Polarizability (BPOL).
\par
In Fig. 3 we report $C(\omega)$ for the DID and BPOL scattering mechanisms for
650$\times$650 2D site percolators at threshold. The interesting features
in
this figure are: (i) The BPOL spectrum shows a single slope, $m = 1.24$,
down to the minimum frequency, (ii) the DID spectrum clearly shows a
crossover at $\omega \approx 8\times10^{-3}$ from $ m \approx 0.94$ (high
frequency)
to $m\approx 0.8$, which was uncovered in \cite{4lettera,5colori} due to
limited size and statistics.
\par
In Fig. 4 we show the same coupling coefficients for 3D site percolators of
linear size
80. In agreement with the results of diagonalization,
\cite{4lettera,30nota1}
there is a very evident crossover in the DID case, while the BPOL
spectrum again is a straight line in almost the whole frequency range. The open
circles
represent the DID spectrum of clusters of linear size 40. Apart from the
obvious fact that the minimum frequency is greater in this case, we do not
observe any significant difference with respect to the larger clusters,
indicating that, at least for $C(\omega)$, finite size effects play a minor
role, if any. It is interesting to mention that removal of dangling bonds
\cite{34nota3} in the site percolators does not change the look and the
slope of the BPOL coupling coefficients both in 2D and 3D, but it does change
the
density
of states which does not appear to follow an $\omega^x$ law in the whole
frequency range; this effect is particularly evident in 3D.
\par
The $C(\omega)$'s relative to 2D (linear dimension 500) and 3D (linear
dimension
70) \cite{33nota2} bond percolators are reported in Figs. 5 and 6
respectively, for both scattering mechanisms DID and BPOL. The BPOL spectra
are well fitted by straight lines with almost the same slopes ($m_{2D} =
1.26$, $m_{3D} =1.58$) as for site percolators. The DID spectra have a
roundish look, but in the low frequency part they are reasonably fitted by
straight lines.
\par
For bond percolators, The nearest-neighbor DID coupling coefficients in 2D and
3D
are practically identical
to DID, in agreement with what found in ref. \cite{6ericstoll}, and we
will not report them.
This fact can be easily understood considering that in bond percolators
(contrary to site percolators) it is possible to find nearest neighbor sites
 which are not connected by any bond. This results in highly
uncorrelated motions of nearby masses that produce most of the DID
scattering.
\par
In Fig. 7 we show the effect of the so called electrical disorder on the DID
scattering of 2D
site percolators. Each mass is randomly assigned one of
two values $\alpha_1$ or $\alpha_2$ of the bare polarizability; the curve
of Fig. 3 ($\alpha_1 = \alpha_2 =1$) is also reported for comparison. Note
that spectra with different electrical disorders coalesce at low
frequency ($\omega < 2 \times 10^{-2}$). Qualitatively the same
result is obtained with all other types of clusters, with and without
dangling bonds. In the case of BPOL for 2D site percolators the $C(\omega)$'s
with
electrical disorder show a change of slope at $\omega \approx 10^{-2}$.
\par
In Fig. 8 are reported the DID and BPOL spectra for 3D site percolators
having a concentration ($c=0.5$) higher than the percolation threshold
concentration, together with the relative density of
states. The latter exhibits the well known \cite{35grest} crossover at
$\omega_{\rho} \approx 7 \times 10^{-2}$ with change of slope. The fact
that neither slope is equal to what expected for phonons ($m=2$) or fractons
($m=0.31$) is also known \cite{35grest}. The $C(\omega)$'s also change
their
frequency dependence at a frequency, $\omega_{C} \approx 6\times 10^{-2}$,
which is apparently
slightly lower than $\omega_{\rho}$. The spectra are rather noisy and it is
impossible to judge whether $C(\omega)$ is a straight line at low frequency,
but in any case
it is clear that below $\omega_{C}  $ the "slopes"
are very different from what was found for the same percolators at
threshold, see Fig. 4. The same qualitative behavior is observed in 2D, and
for bond percolators, though the 3D spectra in this case are even more
noisy at low frequency.
\par
The slopes observed are summarized in Table I for the different kinds of
percolators and scattering mechanisms considered; in the case of DID, where
different slopes ore found in the same spectrum, we report the low frequency
ones
for the reason discussed in the next section.

\section{Discussion}
\label{disc}
{}From the results presented in the previous section, it appears that
$C(\omega)$
behaves quite differently depending on the scattering mechanism. In fact,
in
the case of BPOL the numerically computed coefficients follow very closely
the law $C(\omega) \propto \omega^x$ almost in the whole investigated frequency
range;
the
values of $x$ do not depend appreciably on site- or bond-percolation, nor
on the
presence or absence of dangling bonds. The situation is rather
different for DID scattering; with this
mechanism, in none of the case studied does $C(\omega) \propto \omega^x$  hold
in
the whole range of frequencies. Following the arguments of refs.
\cite{6ericstoll,7orbnippi}, we shall assume that it is the low frequency
part of the DID spectra that is to be compared with the proposed scaling
laws (but it is not clear to us why the arguments of those references
do not apply to BPOL). This is supported by the result of Fig. 7,
that shows  that the effect of
electrical disorder, which is a local random disturbance and as such is not
expected to scale, disappears at more-or-less the same frequency
($\omega \approx 10^{-2}$) where $C(\omega)$ changes slope.
\par
Since the DID and BPOL coupling coefficients  are so different and produce
different slopes,
we can
compare our results only with the theoretical model of Alexander et al
\cite{14physica} because to our knowledge this is the only paper where
different scattering mechanisms are considered. Alexander et al
consider DID and NNDID: the latter coincides
with BPOL for site percolators and, as mentioned, it is practically
identical to DID for bond percolators, so that it is possible to compare our
numerical data with the formulas of ref.\cite{14physica} that are:
\begin{equation}
	C(\omega)_{DID} \propto \omega^{(2 \overline d/D)(\sigma+d)-3\overline d}
\label{al1}
\end{equation}
and
\begin{equation}
	C(\omega)_{NNDID} \propto \omega^{2\overline d \sigma/D}
\label{al2}
\end{equation}
where $\sigma$ is a scaling index introduced by Alexander et al
\cite{14physica} whose value is not known {\it a priori}: for a homogeneous
medium $\sigma=1$, while $\sigma <1$ means a violation of scaling
\cite{14physica}. Values of $\sigma \approx 1$ were deduced in ref.
\cite{7orbnippi} on the basis of simulations \cite{6ericstoll} performed on
smaller clusters than done here. Equation (\ref{al2}) was first
derived by Boukenter et al \cite{12bouk}.
\par
Fitting the slopes with equations (\ref{al1})
and (\ref{al2}) we obtain the values
of $\sigma$ reported in Table I. The picture that emerges is complex.
\begin{itemize}
\item ($i$) In two cases, 3D site percolators (DID) and 2D bond
percolators (DID) we find $\sigma \approx 1$, which is the desired value;
\item ($ii$) in two cases, 2D site percolators (NNDID) and 3D bond
percolators (DID), the
values found are smaller than 1, but not too much ($\sigma \approx 0.9$);
\item ($iii$) in the remaining four cases, $\sigma$ is either too large to be
plausible ($\sigma \approx 1.4$ for 2D
site percolators (DID) and 3D site percolators (NNDID)) or exceedingly
small ($\sigma \approx 0.3$ for 2D and 3D bond percolators (NNDID)).
\end{itemize}
The first thing that we note is that so far $\sigma$ has always
been thought of as a parameter
that characterizes the scattering {\it system}, not the scattering {\it
mechanism}, so that the wide variations observed in Table I between DID and
NNDID apparently imply that equations (\ref{al1})
and/or (\ref{al2}) are not right.
\par
On the other hand, the fact that the BPOL $C(\omega)$'s are straight lines over
a
frequency range
of more than two orders of magnitude, and are so insensitive to the nature of
the
system, might reasonably be taken as
indication
that for this scattering mechanism scaling holds, but
following a law other than equation (\ref{al2}) for site percolators. In our
opinion it would not be too much of a surprise if BPOL should scale and DID
should not because the
modulation of polarization in DID proceeds through electromagnetism, which
definitely does not propagate along the fractal paths. In any case, we
think
that nothing should be taken for granted when treating this subject. For
example, as mentioned in the previous section, removal of dangling bonds
from 2D and (especially) 3D site percolators produces a situation where the
mass distribution follows the same power law as in the case with dangling
bonds, $C_{BPOL}(\omega)$ does the same, but the density of states is no longer
a
power law for $\omega >\approx 4 \times 10^{-2}$. This situation might even be
interpreted as indication that BPOL's being a straight line has actually
nothing to do with dynamic scaling: in fact, in these systems
$C_{BPOL}(\omega)$
is a straight line in a frequency region where the density of states (i.e. the
most believed scaling dynamical quantity) is not. At present we are not able to
resolve this question.
\par
Another result in Table I that is worth a comment is the value of $\sigma$ for
2D and 3D bond percolators under the NNDID scattering mechanism. The fact that
for these percolators DID and NNDID $C(\omega)$'s are almost identical implies
that at all frequencies available to the present simulation
the scattering coefficient is mostly determined by pairs of nearest
neighbor masses that are not connected by a bond. This situation produces
$\sigma \approx 0.3$ for NNDID. In principle, however, it cannot be
excluded that on much larger clusters the contribution of these pairs may
become less important. In order to find $\sigma \approx 1$
\cite{36nota4} the simulation should yield NNDID slopes of $\approx 1.1$ (3D)
and $\approx 1.4$ (2D), i.e. very different from the values obtained here. In
any
case, these new slopes would be found in a frequency range
($\omega \sim < 10^{-4}$) so low to be of little physical significance.
\par
This remark introduces another important issue, i.e. the possibility of
extracting from
experimental spectra information on the static and dynamic fractal
parameters. This is in general done by fitting the low-frequency parts of
the spectra (where scaling is expected to hold) with expressions like Eqs.
(\ref{al1}) or (\ref{al2}). Assuming that one has the right equations, one
obvious
shortcoming
of this procedure is that one should know the scattering {\it mechanism}
(DID, BPOL, NNDID, a mixture, ...), which has been shown to be so important in
determining
the low
frequency slope. But even if this was known, there is another problem
that
is evident from fig. 8. In fact, real disordered
materials cannot reasonably be thought of as percolators at threshold: if
the percolation model is to have a sense, it must be a percolator above
threshold. From fig. 8 we see that in this case the values of the "slopes"
bear no resemblance with those found at threshold, and especially so at low
frequency. This fact has long been known as regards the density of states
of
percolators above threshold \cite{35grest}. Of course, the particular
concentration value used here ($c$=0.5) has no special meaning, nor is
it thought to represent any particular physical system.
\par
In conclusion, we think that the search for the scaling laws which might
possibly govern Raman scattering from fractal objects is interesting in
itself and as such worth being pursued; on the basis of the present
numerical simulations, we think that these laws have not been found yet.
Even more disputable is the extraction of fractal parameters from the
experimental Raman spectra. In this regard, we frankly hope that the
present
paper may encourage a realistic reconsideration of previous work, even if
this might lower the role of the fractal model of disordered solids.

\acknowledgements
We are indebted to M. Sampoli and L. Sampoli for  useful discussions
and for bringing to our attention
the similarities between the  method of moments and the Lanczos procedure.

\appendix

\section*{Definition of the initial vectors}
The general expression for the Raman intensity in
$\alpha,\beta$ polarization at energy $\hbar \omega$ and exchanged momentum
$\hbar {\bf k} $ is given by:
\begin{equation}
I_{\alpha\beta}(\omega) = {\cal A} \int \; dt \; e^{i\omega t} \sum_{ll'}
\langle \pi^l_{\alpha\beta}(t) \pi^{l'}_{\alpha\beta}(0)
\; e^{i {\bf k} \cdot ({\bf R}_l(t) - {\bf R}_{l'}(0) ) } \rangle
\label{e1}
\end{equation}
where $\cal A$ is a constant and $\pi^l_{\alpha\beta}(t)$  is the
 $\alpha,\beta$ component of the effective polarizability tensor
 of the $l$-th atom placed at site
  ${\bf R}_l(t)$. The time dependence of this tensor is caused by
 the relative displacement  of atom $l$ with respect to all  other atoms.
\par
The instantaneous position ${\bf R}_l(t)$ may be written as
$$
{\bf R}_l(t) = {\bf x}_l + {\bf u}_l(t)
$$
where ${\bf u}_l(t)$ is the displacement from the equilibrium position
${\bf x}_l$.
For small displacements, both the effective polarizability and the
exponential function in equation (\ref{e1}) can be expanded in power series:
\begin{equation}
e^{i{\bf k} \cdot {\bf R}_l(t)} \approx e^{i {\bf k} \cdot {\bf x}_l} \
(1 + i {\bf k} \cdot {\bf u}_l(t))
\label{e2}
\end{equation}
$$
\pi^l_{\alpha\beta}(t) \approx \pi^l_{\alpha\beta} +
\sum_m \sum_\gamma \; \;
\frac{\partial \pi^l_{\alpha\beta}}{\partial u_{m \gamma}}
\; \; u_{m \gamma}(t).
$$
The displacement, in turn, may be expanded in normal modes:
\begin{equation}
u_{m\gamma}(t) \; = \; \sqrt{\frac{\hbar}{2 M N}} \sum_\lambda
 \; \frac{1}{\sqrt{\omega_\lambda}} e_{m\gamma}(\lambda) A_\lambda(t)
\label{e3}
\end{equation}
where $A_\lambda(t)$ are the normal  coordinates that (for the Stokes
part of the spectrum) obey the relation:
\begin{equation}
\int \; dt \; e^{i \omega t} \langle A_\lambda(t) A_{\lambda'}(0) \rangle =
\delta_{\lambda\lambda'} \frac{[n(\omega)+1]}{\omega}
\delta(\omega-\omega_\lambda)
\label{e4}
\end{equation}
{}From the previous equation we have
\begin{equation}
I(\omega)= {\cal A} \frac{[n(\omega)+1]}{\omega} \rho(\omega) C(\omega)
\end{equation}
having defined
\begin{equation}
C(\omega)=\sum_\lambda \vert C_\lambda \vert^2 \delta(\omega-\omega_\lambda) /
\sum_\lambda \delta(\omega-\omega_\lambda)
\label{g1}
\end{equation}
$$
=\sum_\lambda \vert C_\lambda \vert^2 \delta(\omega-\omega_\lambda) /
\rho(\omega)
$$
and
\begin{equation}
C_\lambda \; = \; \frac{1}{N} \sum_{ml} \sum_\gamma
\frac{\partial \pi^l_{xy}}{\partial u_{m \gamma}}
[e_{l\gamma}(\lambda)-e_{m\gamma}(\lambda)].
\label{f1}
\end{equation}
\par
In writing the above equations we have assumed that we are interested in
the depolarized component of the spectrum, $(\alpha,\beta)=(x,y)$.
Considering that
$$
\sum_{m} \frac{\partial \pi^l_{xy}}{\partial u_{m \gamma}}=0
$$
equation (\ref{f1}) may be cast in the form
\begin{equation}
C_\lambda \; = \; \frac{1}{N} \sum_{ml} \sum_\gamma
\frac{\partial \pi^l_{xy}}{\partial u_{m \gamma}}
e_{m\gamma}(\lambda).
\label{v1}
\end{equation}
or
\begin{equation}
C_\lambda=\sum_m \sum_\gamma p_{m\gamma} e_{m\gamma}(\lambda)
\end{equation}
where
\begin{equation}
p_{m\gamma}= \; = \; \frac{1}{N} \sum_{l}
\frac{\partial \pi^l_{xy}}{\partial u_{m \gamma}}.
\label{f2}
\end{equation}
$C_\lambda$ in equation (\ref{v1}) has the same structure as $d_\lambda$
in equation (\ref{dl}), so that the components of the initial vector
${\bf t}^{(0)}$ are $p_i \equiv p_{m\gamma}$.
We note that from equation (\ref{g1}) we obtain
\begin{equation}
\frac{\rho(\omega) \ C(\omega)}{2 \omega} =
\sum_\lambda \vert C_\lambda \vert^2
\frac{\delta(\omega-\omega_\lambda)}{2 \omega}=
\sum_\lambda \vert C_\lambda \vert^2 \delta(\omega^2-\omega^2_\lambda)
\end{equation}
to be compared with equation (\ref{fo2}) in the text.
\par
In this paper we have considered different mechanisms of polarizability
modulation, i.e. different expressions for $\pi_{\alpha\beta}^l$, that
result in different ${\bf t}^{(0)}$'s. In particular, the explicit
expressions for the studied scattering mechanisms are:
$$
{\rm DID:}\;\;\;\;\;
\pi_{\alpha\beta}^l \ = \ \sum_m {\bf T}^{(2)}_{\alpha\beta}(lm)
\alpha_l \alpha_m
$$
$$
{\rm NNDID:}\;\;\;\;\;
\pi_{\alpha\beta}^l \ = \ \sum_{m ,\{l\}}
{\bf T}^{(2)}_{\alpha\beta}(lm) \alpha_l \alpha_m
$$
$$
{\rm BPOL:}\;\;\;\;\;
\pi_{\alpha\beta}^l \ = \ \sum_{m, \{l\}}
{\bf T}^{(2)}_{\alpha\beta}(lm) V_{lm}
$$
where $\alpha_l$ is the bare polarizability of atom $l$,
${\bf T}^{(2)}_{\alpha\beta}({\bf r})$ is the dipole propagator
$$
{\bf T}^{(n)}_{\alpha_1 \alpha_2 .. \alpha_n}({\bf r}) =
\nabla_1 \nabla_2 ... \nabla_n \frac{1}{\vert {\bf r} \vert}
$$
the index $m,\{l\}$ indicates that the summation
concerns all the  $m$ nearest neighbors of atom $l$, and
$V_{lm}$ is 1 or 0 according to whether atoms $l$ e $m$
are connected by a bond.
\par
Therefore, from equation (\ref{f2}) we find that:
$$
{\rm DID:}\;\;\;\;\;
p_i \ = \ \sum_l
{\bf T}^{(3)}_{xy\gamma}(lm)
\alpha_l \alpha_m
$$
$$
{\rm NNDID:}\;\;\;\;\;
p_i \ = \ \sum_{l ,\{m\}}
{\bf T}^{(3)}_{xy\gamma}(lm) \alpha_l \alpha_m
$$
$$
{\rm BPOL:}\;\;\;\;\;
p_i \ = \ \sum_{l, \{m\}}
{\bf T}^{(3)}_{xy\gamma}(lm) V_{lm}
$$
\par
In the case of DID the effective polarizability depends on the positions
of all atoms, each weighted by the factor $1/R^3$; in the NNDID case this
effect is limited to nearest neighbors, to simulate induction mechanisms
that decay faster than $1/R^3$. In the case of BPOL modulation of
polarizability occurs only if the nearest neighbors are actually connected
by a bond. Therefore, in site percolators NNDID and BPOL coincide, while
they do not in bond percolators.
\par
We have also been interested in the calculation of the density of states;
in this case the values of $p_i$ are
uniformly and randomly distributed between -0.5 and 0.5, as shown in ref.
\cite{1benoit1}.


\newpage
\begin{figure}
\caption{Log-log density of states of 2D site percolators (average of
10 realizations) having linear dimension $L$=650, with $\varepsilon=2 \cdot
10^{-6}$ and: (a) 5000 moments; (b) 1000 moments; (c) 50 moments. $\omega =
\omega/\omega_{max}$; $m$ is the fitted slope; the curves are shifted
vertically
for graphical convenience.}
\end{figure}

\begin{figure}
\caption{Scalar product $({\bf t}^{(0)}, {\bf t}^{(n)})$ (see text) as a
function of the number of iterations
$n$ for ordered square lattices of linear sizes 50 (a)
and 100 (b), and 2D site percolators at threshold of linear sizes 650 (c)
and 50 (d).}
\end{figure}

\begin{figure}
\caption{$C(\omega)$ for 2D site percolators, $L$=650. (a) DID; (b) BPOL.
Average of 10 realizations.}
\end{figure}

\begin{figure}
\caption{$C(\omega)$ for 3D site percolators, $L$=80. (a) DID; (b) BPOL.
Average of 20 realizations.  Open circles: DID for 3D site
percolators, $L$=40, average of 100 realizations.}
\end{figure}

\begin{figure}
\caption{$C(\omega)$ for 2D bond percolators, $L$=500. (a) DID; (b) BPOL.
Average of 10 realizations.}
\end{figure}

\begin{figure}
\caption{$C(\omega)$ for 3D bond percolators, $L$=70. (a) DID; (b) BPOL.
Average of 10 realizations.}
\end{figure}

\begin{figure}
\caption{Effect of electrical disorder on $C(\omega)$ for 2D site
percolators, $L$=650. (a) $\alpha_1$=0, $\alpha_2$=2; (b)  $\alpha_1$=0.5,
$\alpha_2$=1.5; (c) $\alpha_1$=$\alpha_2$=1.  Average of 10
realizations.}
\end{figure}

\begin{figure}

\caption{$C(\omega)$ for 3D site percolators, $L$=70, above percolation
threshold ($c$=0.5). (a) Density of states; (b) DID; (c) BPOL;
the curves are vertically shifted for graphical convenience. Average of
10 realizations}
\end{figure}

\begin{table*}
\caption
{
Fitted slopes, $m$,  for $C(\omega)$, see Figures (3) to (8),
and  values of $\sigma$ obtained by fitting the numerical data
		with equations (15) and (16) for DID and NNDID
		respectively. For BPOL no such theoretical expression is available,
	but in the case of site percolators BPOL coincides with NNDID.
}
\begin{tabular}{ccccccc}
&&\multicolumn{2}{c}{DID} & \multicolumn{2}{c}{NNDID} &
{BPOL} \\
\hline

&& $m$ & $\sigma$& $m$ &$\sigma$&
$m$ \\
\hline\\
site    &       2D      & $0.80 \pm 0.02$ & $1.40 \pm 0.01$ & $1.24 \pm 0.01$&
$0.89 \pm 0.01$ & $1.24 \pm 0.01$ \\
percolator&     3D      & $0.41 \pm 0.03$ & $0.97 \pm 0.03$ & $1.56 \pm
0.01$& $1.40 \pm 0.01$ & $1.56 \pm 0.01$ \\
\hline\\
bond    &       2D      & $0.40 \pm 0.05$ & $1.12 \pm 0.04$ & $0.40 \pm 0.05$
& $0.30 \pm 0.05$ &$1.26 \pm 0.01 $\\
percolator&     3D      & $0.30 \pm 0.08$ & $0.86 \pm 0.07$ & $0.30 \pm 0.08$
&$0.30 \pm 0.05$  & $1.58 \pm 0.01 $\\
\end{tabular}
\end{table*}


\begin{thebibliography}{99}
\bibitem{10minfreq} The average minimum frequency of the percolators scales
as $\omega_{min}(L) = c L^{-D/\overline d}$; the constant $c$ was determined by
averaging the minimum eigenvalues obtained by diagonalizing the dynamical
matrix of clusters with $L=30,40,50$ in 2D, and $L=10,15,20$ in 3D.  In this
way we obtain the values $c_2=4.5 \pm 0.2$ in 2D and $c_3=3.6 \pm 0.1$ in 3D.
These values of $c$ result in the following ratios
$R_d(L)=\omega_{min}/\omega_{max}$:
$R_2(650) \approx 1.6 \times 10^{-4}$ in 2D, and $R_3(80) \approx 5\times
10^{-4}$ in 3D.  However, in the first decade the levels are so separate in
a log-log plot that it is meaningless to speak about the density of states.
It is interesting to follow the behavior of the spectrum in this frequency
range with higher resolution than in Fig. 1 ($\varepsilon = 10^{-9}$) and
with different numbers of moments.  The lowest
frequency peak in the density of states shifts to low frequency by more
than
one order of magnitude as the number of moments is increased from 1000 to
10,000, and at the same time the number of peaks in the reduced frequency
decade $10^{-4}\div 10^{-3}$ increases from 1 to about 10.  The spectrum
becomes smooth in the next decade.
\bibitem{33nota2} The calculation for bond percolators requires more
computer memory, and this is the reason for the smaller linear dimensions.
\bibitem{34nota3} It should be stressed that removal of dangling bonds does
 not produce what is known as the backbone of the percolator.
\bibitem{36nota4} According to the analysis of ref.  \cite{14physica},
though
$\sigma >1$ is consistent with scaling, a "strong self similarity
assumption
would really imply $\sigma=1$ even for a fractal".
\end{thebibliography}
\end{document}